# A LabVIEW based user-friendly X-ray phase-contrast imaging system software platform

Shenghao Wang[1], Huajie Han[1], Kun Gao[1], Zhili Wang[1], Can Zhang[1], Meng Yang[1], Zhao Wu[1], Augusto Marcelli[2, 1], Ziyu Wu[1,*]

[1]National Synchrotron Radiation Laboratory, University of Science and Technology of China, Hefei 230027, China

[2]Istituto Nazionale di Fisica Nucleare, Laboratori Nazionali di Frascati, P.O.Box13, 00044 Frascati (RM), Italy

*Corresponding author: wuzy@ustc.edu.cn, Phone: +8655163602077, Fax: +8655165141078

**Abstract:** X-ray phase-contrast imaging can provide greatly improved contrast over conventional absorption-based imaging for weakly absorbing samples, such as biological soft tissues and fibre composites. In this manuscript, we introduce an easy and fast way to develop a user-friendly software platform dedicated to the new grating-based X-ray phase-contrast imaging setup recently built at the National Synchrotron Radiation Laboratory of the University of Science and Technology of China. Unified management and control of 21 motorized positioning stages, of an ultra-precision piezoelectric translation stage and of the X-ray tube are achieved with this platform. The software package also covers the automatic image acquisition of the phase-stepping scanning with a flat panel detector. Moreover, a data post-processing module for signals retrieval and other custom features are in principle available. With a seamless integration of all necessary functions in a unique package, this software platform will greatly support the user activity during experimental runs.

1. **Introduction:**

Since Wilhelm Conrad Röntgen discovered X-ray radiation in 1895, X-ray imaging has found wide applications in medical imaging, social security, inspection tests, materials science, and many industrial fields. However, for soft tissues, such as adipose tissues, muscles, connective tissues or cartilages it has a long-standing reputation of a low contrast resolution. This is the main reason that ruled out X-ray computed tomography for the investigation of musculotendinous trauma, infections, neoplasia and many other medical diagnosis [1].

To overcome the limitation, X-ray phase-contrast imaging was proposed as it could provide greatly improved contrast over conventional absorption-based imaging in biological samples, polymers or fibre composites [2-4]. The demonstration of a Talbot-Lau interferometer in the hard X-ray region with a conventional low-brilliance X-ray source overcome the problems that impaired a wider use of phase-contrast in X-ray radiography and tomography, representing a great breakthrough in X-ray imaging [5]. It also foreseen many potential applications in biology, medicine, non-destructive testing, food inspection, security devices, etc. [6-9].

A new grating based X-ray phase-contrast imaging system has been designed and assembled at the National Synchrotron Radiation Laboratory of the University of Science and Technology of China.

This high-grade precision and sophisticated imaging system is mainly made by a X-ray tube, a X-ray flat panel detector and three micro-structured gratings, all mounted on multi-dimensional motorized optical displacement tables assembled by 21 motorized positioning stages and an ultra-precision piezoelectric translation stage. The stages are located inside a radiation protection lead room, and can only be manually controlled when the X-ray generator is shut off. However, with this setup, tedious movements of the stages are needed during experimental runs for the grating alignment with X-ray radiation before the image acquisition. It is then urgent the development of a remote control package for all the motorized stages. Also an automatic image acquisition with the flat panel detector is required, because in phase-stepping scanning procedure, dozens of images have to collected to generate a single phase-contrast projection image and hundreds of projections are needed in the computed tomography for a high-quality 3D reconstruction [10]. Moreover, once we obtain raw images series, a complex digital image process has to be performed to retrieve conventional absorption, phase-contrast and dark field signals, respectively.

Typically, the LabVIEW protocol is used to control motorized stages and X-ray tubes thanks to the easy availability of hardware drivers [11, 12]. Similarly, $C^{++}$ is utilized to develop a customized image acquisition because the software development kit (SDK) provided by the flat panel detector supplier comprises a set of standard C-functions [13], while digital image processing is usually performed by MATLAB [14]. Utilization of different program frameworks in an experiment greatly decrease the maneuverability of an imaging system, especially for users with a limited programming experience in text based languages.

In order to overcome the aforementioned problems, we intend to develop a software platform, where all necessary functions of experimental runs are included, and it has also to be user-friendly and easy to learn for new users. As a graphical programming language, LabVIEW is a popular development tool used both in scientific researches and in industrial applications, which has already applied in building traditional absorption-based X-ray imaging systems [15-17]. In this manuscript, we first describe the layout and the working principle of the new grating based X-ray phase-contrast imaging setup, then we will describe in details how to deal the instruments control, the image acquisition and the data post-processing using LabVIEW. Finally, remarkably improved X-ray phase-contrast imaging results of photoresist SU8 will be presented.

2. **System setup and working principle:**

Fig. 1 is the mechanical structure of the grating based X-ray phase-contrast imaging setup, built from commercial available components and other custom made devices. It is mainly composed by an X-ray tube, an X-ray flat panel detector and three micro-structured gratings. These components are assembled on multi-dimensional motorized optical displacement tables moved by 21 motorized positioning stages (15 translation stages, 3 rotary stages and 3 goniometric stages). Because of the demanding precision, an ultra-precision piezoelectric translation stage with an encoder has been used for the phase stepping scanning.

Working principles of this X-ray phase-contrast imaging system is described below. The source grating G0 is placed close to the X-ray tube anode, creating an array of individually coherent, but mutually incoherent line sources. The grating G1 acts as a phase mask, generating periodic phase modulations of the x-ray wave front. Through the self-imaging phenomenon, the phase modulation is transformed into an intensity modulation in the plane of the grating G2, forming a linear periodic fringe pattern perpendicular to the optical axis and parallel to the grid of G1. The second grating G2 with absorbing lines, is placed at the position and with the same period and orientation of the fringes generated by G1. The X-ray flat panel detector, sets close to the grating G2, is used to collect the image.

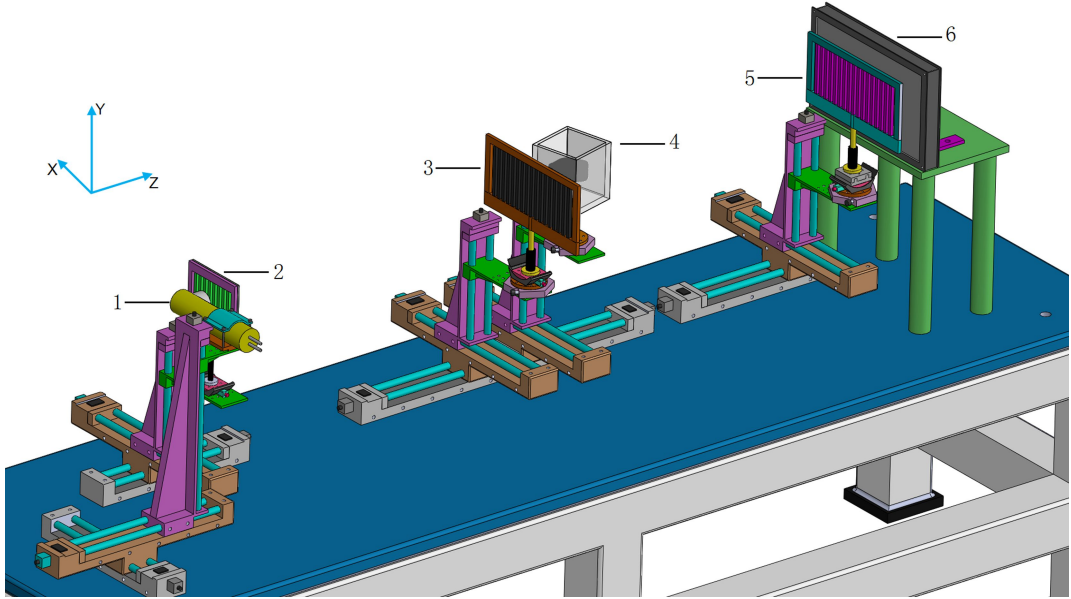

Figure 1: Mechanical layout of the grating based X-ray phase-contrast imaging setup. 1. X-ray tube, 2. Source grating G0, 3. Beam-splitter grating G1, 4. Sample room, 5. Analyzer grating G2, 6. Flat panel detector. For the precise alignment of the system, the following movements are motorized: X-ray tube: translation (along x-, y-, and z-axis); G0 grating: translation (along x-, y- and z-axis), tilt (around z-direction); G1 grating: translation (in x-, y-, and z-axis), rotation (around y-axis), tilt (around z-axis); Sample stage: translation (along x-, y- and z-axis), rotation (around y-axis); G2 grating: translation (along x-, y-, and z-axis), rotation (around y-axis), tilt (around x-axis), high precision translation (along x-axis).

To ensure that the Moiré fringe generated by each line source of the grating G0 may overlap incoherently at the detector plane, the geometry of this setup has to satisfy the condition.

$$\frac{p_0}{l} = \frac{p_2}{d} \quad (1.1)$$

where $p_0$, $p_2$ are the periods of grating G0 and G2, respectively, $l$ is the distance between grating G0 and G1, and $d$ represents the distance between the gratings G1 and G2.

The differential phase-contrast image information process achieved by the two gratings G1 and G2, essentially relies on the fact that the sample placed in the X-ray beam path induces a slight refraction of the beam transmitted through the object. The fundamental idea of the differential phase-contrast imaging depends on locally detection these angular deviations. The angle $\alpha$ is proportional to the local gradient of the object's phase shift, and it can be written as

$$\alpha = \frac{\lambda}{2\pi} \frac{\partial \Phi(x,y)}{\partial x} \quad (1.2)$$

Where $\Phi(x,y)$ is the phase shift of the wave front and $\lambda$ is wavelength of the radiation.

Determination of the refraction angle can be achieved by phase stepping [18], a typical measurement strategy, which is based on a set of images taken at different positions of the grating G2. When G2 is scanned along the transverse direction, the signal intensity in each pixel of the detector oscillates as a function of the grating position. By a Fourier analysis of the intensity curve, the conventional absorption projection of the sample and the refraction signal can be simultaneously retrieved for each pixel. Moreover, a dark-field image can also be acquired at the same time [19].

## 3. LabVIEW-based system software platform:

### 3.1 Instruments control framework:

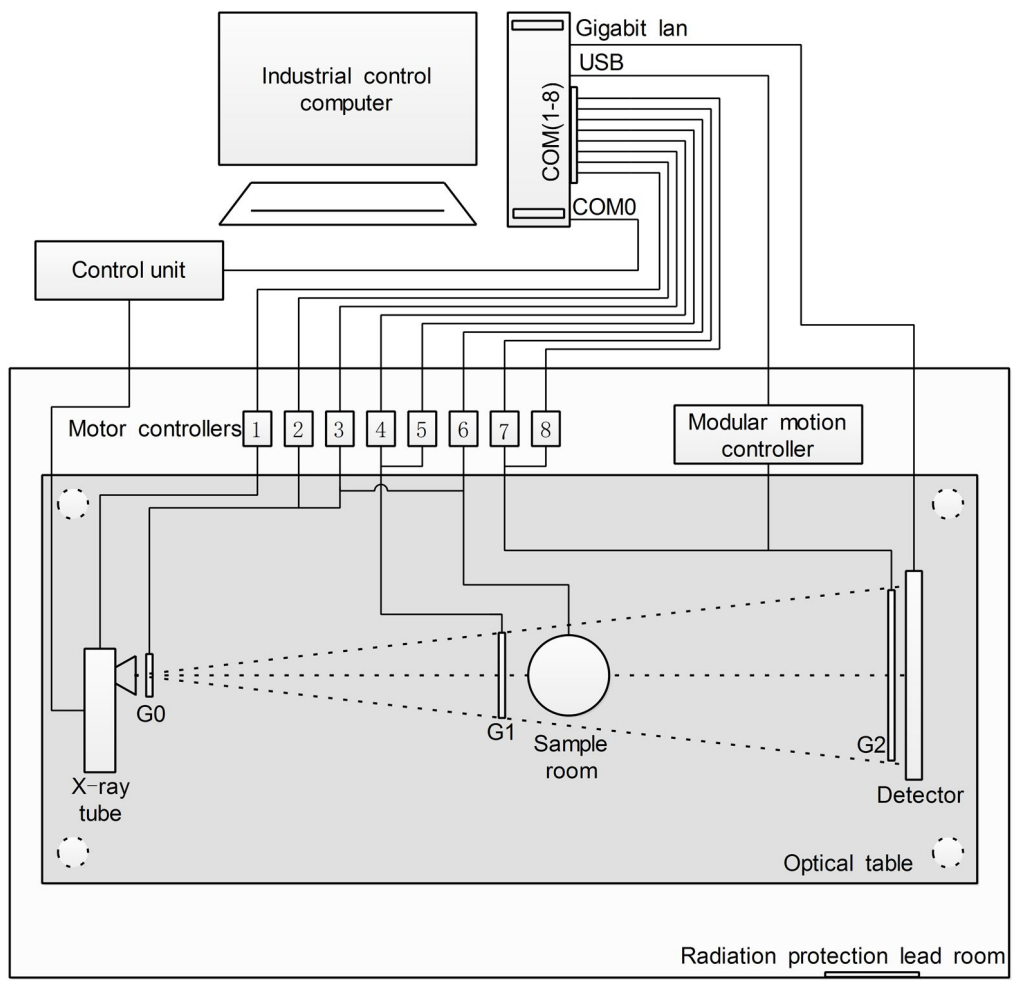

Figure 2: The control framework of the grating based X-ray phase-contrast imaging system.

Communications of all instruments with a custom-made industrial control computer have been realized in different ways following correlated protocols, as schematically illustrated in Fig. 2. The X-ray tube is controlled by its control unit, which manually operates the X-ray source. In order to control it with an external PC, the control unit is connected using the serial port COM1. The 21 precision motorized stages are controlled by eight stepper motor controllers. Each of them may load up to three stages and the communication between each motor controller and PC is realized by one serial port. The eight serial ports (COM1-COM8) are drawn forth by a serial-extended card installed on the motherboard. The USB port is utilized here to connect the modular motion controller and the PC. In this way, data transmission is achieved between PC and piezoelectric stages. The Gigabit Ethernet Interface is embedded in the X-ray flat panel detector and is able to interact with the standard Gigabit Ethernet network interfaces (RJ45) for data transmission between the X-ray flat panel detector and the industrial control computer. The interface is used in direct (Point-to-Point) connection to the PC to guarantee that no additional network traffic arises for this link.

### 3.2 Program development strategy:

The software development is mainly made of three function modules, i.e., the instrumental control of the motorized stages and the X-ray tube, the image acquisition with the flat panel detector and the data post-processing for the signal retrieval. The program development strategy of each module is described in the following.

#### 3.2.1 Motors & X-ray tube control

For the 21 motorized stages, data communication has been realized with the RS232 protocol. NI-VISA (National Instruments-The Virtual Instrument Software Architecture), as a standard for configuring, programming and troubleshooting instrumentation systems comprising GPIB, VXI, PXI, Serial, Ethernet and USB interfaces, provides here the programming interface between motors and LabVIEW. The motors supplier (Beijing Optical Century Instrument Co., Ltd, China) offers a set of commands for the secondary development. Table 1 is the frequently-used function description and related commands. By sending instructions to the controller and receiving data from it, the corresponding motor stage recognizes common functions, such as relative/absolute movements, velocity setting, position query, emergency stop, zero position, move to positive/negative limit, etc... Sub VI of each function can be developed, and then based on these sub VIs, the desired management of the 21 stages is achieved. In the case of X-ray tube, a similar procedure can be setup to achieve its full control.

Table 1. Description of the frequently-used function and of the corresponding command of the motorized stages, X axis is used here as an example.

| Function description | Command | Function description | Command |
| --- | --- | --- | --- |
| Connect | ?R/ | Move relative | X:number/ |
| Position query | ?X/ | Move to negative limit | -HMX/ |
| Velocity query | ?VX/ | Move to positive limit | HMX/ |
| Setting velocity | VX=number/ | Zero position | HMX/ |
| Move absolute | X=number/ | Emergency stop | S0/ or SX/ |

When dealing with the ultra-precision piezoelectric translation stage, issues are easier to manage because a set of sub VIs are provided by the supplier (Micronix Co., Ltd, Irvine, California USA). We just needed to assemble the sub VIs to realize the customized control. The available sub VIs eliminate the need to learn low-level communication commands, remarkably reducing the development cycle. As a matter of fact, the Instrument Driver Network (NI-IDNet) offers more than 10,000 free drivers for third-party instruments from more than 350 vendors. When users connect an instrument with PC, LabVIEW wizards can automatically download the corresponding free codes. Customized applications with minor changes based on existing VIs can be quickly created.

### 3.2.2 Image acquisition

Things are more complicated to develop in LabVIEW an image acquisition function module for the X-ray flat panel detector (Perkin Elmer Inc., Waltham, Massachusetts, USA). The company supplies the SDK and the X-Ray Imaging Software Library (XISL) of the flat-panel detector. The SDK comprises a selection of Standard C-functions to allow the user to put into action the desired mode of operation. the C-functions in the SDK can be repacked as hardware initialization, parameter configuration, image acquisition, and others, each classification can be encapsulated again into DLL(dynamic link library), by calling this DLL in LabVIEW, desired management of the detector is realized, more details about the development procedure can be referred with this technical report [20].

### 3.2.3 Data post-processing

After the collection of raw images, a complex digital image processing and tedious numeric computations are needed to retrieve conventional absorption, phase-contrast and dark field signal. Within LabVIEW, the data post-processing function module is developed using the NI-VDM toolbox, which contains a complete set of digital image processing algorithms improving the efficiency of the projects and reducing the programming effort of users.

## 3.3 GUI of the software

Fig. 3 is the main GUI of the system software platform. After execution of the software, only the main menu and a system introduction page appear. Users need log in with authorized username and password from the main menu, where "help" and "about" are also available. The introduction page disappears and the main GUI emerges after a successful login. The main GUI is mainly made up of a three-page Tap Control. Fig. 3(a) shows the GUI of the instruments control and the image acquisition. The upper left is the area where is possible to manage the 22 motorized stages, where each stage can be selected by combined assignment of "Device number", "Motor type" and "Axis" menu ring. Relative or absolute motion modes can be set with the "Move mode" menu ring, and velocity and distance units can be also set by other menu ring. Values can be inserted in the "Displacement" numeric control, while the "Position" numeric indicator displays the calculated position of the selected stage by a motor controller based on the pulse numbers. The "Encoder" is only enabled when the piezoelectric stage is chosen, showing the position given by the encoder attached on the piezo stage. The bottom left controls the X-ray tube operation where users may set the acceleration voltage and the current of the tube. Other functions can be also integrated in this area. The image control sets in center of the GUI shows the X-ray image acquired with the flat panel detector. The detector needs to be first initialized and later, users have to set the other parameters such as frame frequency, binning mode, etc. Offset and gain correction of the original data are performed here to subtract the dark current and even out the differences in the gray value, respectively. Live mode of the detector is often used during the gratings alignment procedure, and in this mode the image sequence coming from the camera is relay straight to the display area. While in the integration mode, images are aggregated and the total is divided by the number of frames, multiple images are averaged in order to reduce the statistical noise. Brightness and contrast of the image can be manually adjusted for better visual appearance, alternatively, in a new window, the best visual appearance in a region of interest (ROI) can be automatically achieved.

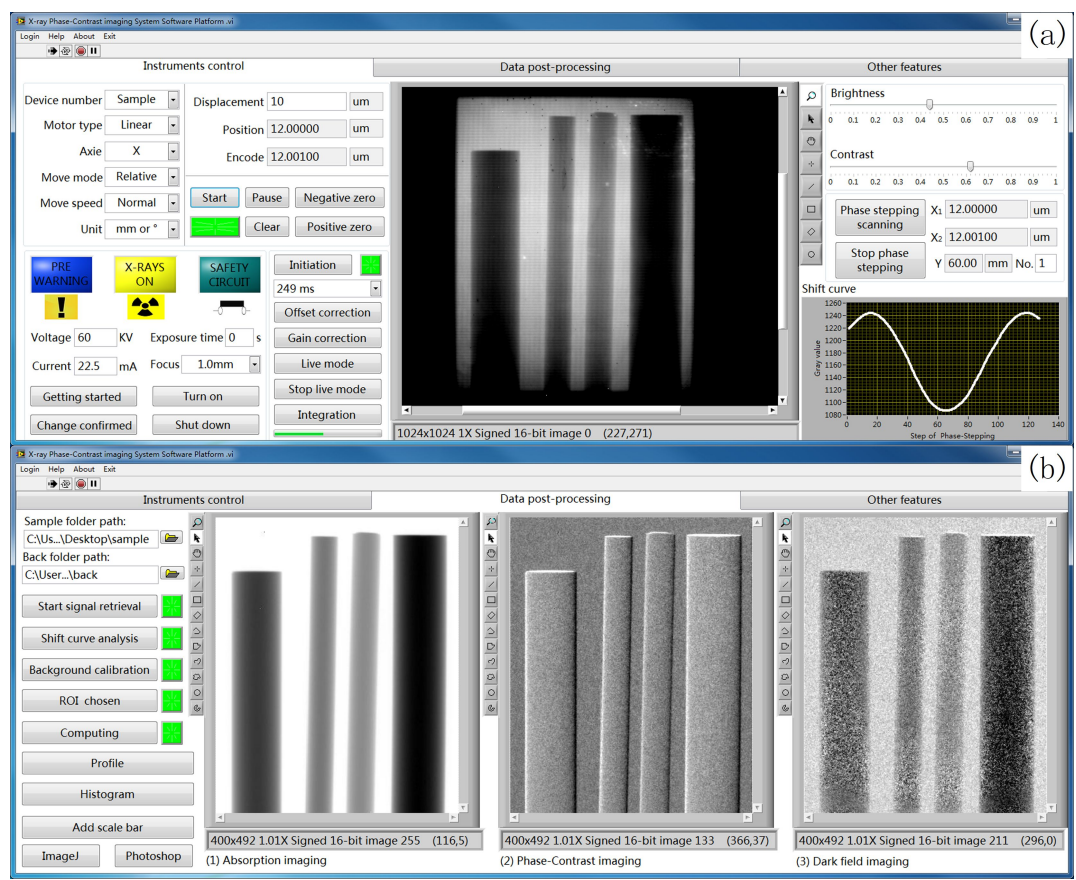

Figure 3: Main GUI of the X-ray phase-contrast imaging system software platform, (a) GUI of the instruments control, and (b) GUI of the data post-processing.

The bottom right part of Fig. 3(a) illustrates the phase stepping scanning module. After the execution, parameters such as steps, exposure time and the number of images to average need to be input in a dialog box, and then, the piezo stage, the image acquisition and the images storage will automatically start. Meanwhile, the real time shift curve of the phase stepping would be generated and appear in a diagram, based on which will users evaluate the scan and judge the time to terminate the experiment. Two phase stepping scanning modes are available, as shown in Fig. 4. The mode (a) is usually applied when the sample under analysis, e.g. a mouse, cannot be wholly fixed on the sample stage, while the scanning mode (b) is used when the sample is absolutely fixed. Compared with the scanning mode (a), although time-consuming, the mode (b) could successfully eliminates negative influence of the stage's return error upon the imaging quality, which is very important in quantitative X-ray phase contrast imaging.

Fig. 3(b) shows the GUI of the data post-processing module. After a successful phase stepping scan and all raw images are stored in the hard disk, the signal retrieval may start. At first, shift curves of the dataset with and without the sample are generated and analyzed so that the starting point, the period of the phase stepping and the fringe visibility can be computed. After, the instability of the X-ray tube and of the detector are carefully calibrated. And then, a custom ROI for signal retrieval is chosen both to decrease the used memory and to minimize the time required in the following procedure. Finally, for each pixel, by a Fourier analysis the sample's conventional attention contrast, differential phase-contrast and dark field signal are retrieved simultaneously, and presented in three image display indicators with an optimized visual appearance. Further processing such as the plotting profile, overlaying ruler, etc. are also available.

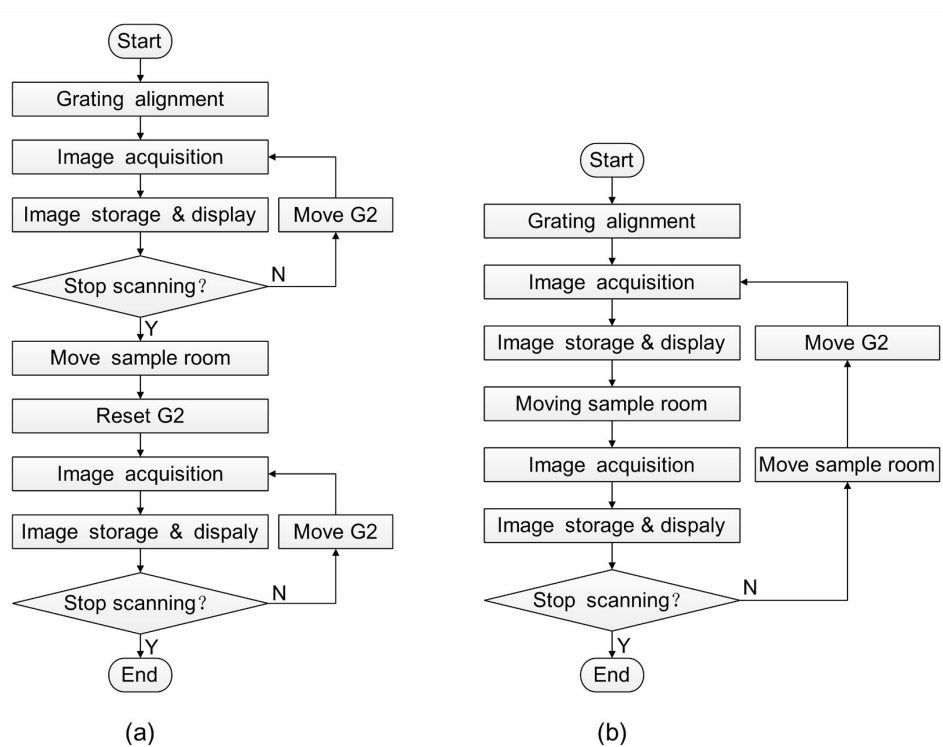

Figure 4: Flow chart of the two phase-stepping scanning mode.

With the "other features" page in the main GUI, an automatic record of the operation history with the imaging system is available, users may also write down experimental notes. In addition, an instruments maintenance function module is also available here.

4. **Experimental result:**

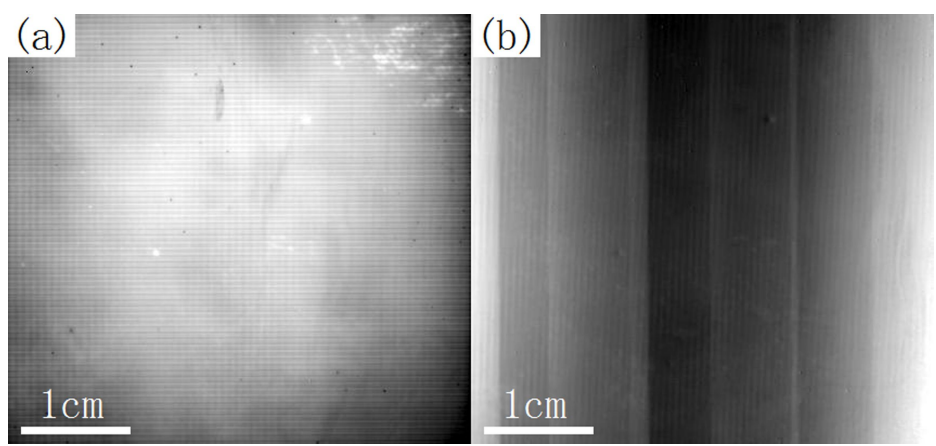

Figure 5: X-ray imaging results of the SU8 photoresist. (a) Conventional X-ray absorption image. (b) Differential phase-contrast image. Images are displayed on a linear gray scale and are windowed for optimized appearance

Fig. 5 illustrates an experiment result obtained with the grating based X-ray phase-contrast imaging system and the software platform above described. The experiment was performed with 45 kV X-ray tube acceleration voltage and a current of 22.5 mA. 50 steps were adopted during the phase stepping scanning procedure, and for each step, 30 raw images were captured to reduce statistical and systemic noise. The images were obtained without an extra sample in the beam path, thus we can consider it as the image of the photoresist (SU8) and the silicon wafer, which support the Au grids. Actually, the Au grids act as an important medium in the signal retrieval, while both photoresist and silicon wafer can be regarded as the sample under analysis. The conventional X-ray absorption image is shown in Fig. 5(a), whereas Fig. 5(b) is the corresponding differential phase-contrast signal. Both images are shown on a linear gray scale and are windowed for best visual appearance. We can see clear vertical stripes in the phase-contrast image while blur dominates in the absorption signal. The vertical strips could be regarded as the uneven photoresist resulting from the ultraviolet exposure technological process during the grating fabrication. Imperfections of gratings in the upper right part of the view are clearly visible in the conventional absorption imaging. On the contrary, we can hardly see them in the differential phase-contrast imaging. This experimental result demonstrates the remarkably enhanced X-ray phase-contrast imaging performance for the SU8 photoresist, also absorption imaging's strong discriminability for imperfections of gratings. By combining attenuation-contrast and phase-contrast signal, significantly more and unique information than any of the techniques alone would be provided by this grating based X-ray phase contrast imaging system.

5. **Discussion and Conclusion:**

In this manuscript we present a user-friendly system software platform based on LabVIEW designed for the new grating-based X-ray phase-contrast imaging system at the National Synchrotron Radiation Laboratory of the University of Science and Technology of China. As a graphical programming language, the easy availability of hardware drivers for a great number of instruments, the easy-to-use multithreaded programming, the efficient graphic user interface design, the high-efficiency debugging functions and many other remarkable features make LabVIEW the ideal software development framework for this imaging setup. Utilization of LabVIEW also greatly decreases the complexity of the problem and shorten the software development cycle.

In the future, with improved gratings, tomographic phase-contrast imaging would be also feasible. Different tomographic phase-contrast scanning mode have been introduced [21-23], all are more complex than traditional absorption based tomographic imaging methods because projection images have to be captured with combined motion of sample' rotation and grating's lateral shift. Function module for phase-contrast tomographic scan would get developed on the basis of the existing software and 3D reconstruction package for absorption, phase-contrast and dark field signal from the same dataset would also be introduced and integrated within this developed platform.

**Acknowledgements:**

The authors gratefully acknowledge Gang Liu and Shuangyue Hou for the grating fabrication. This work was partly supported by the National Basic Research Program of China (2012CB825800), the Science Fund for Creative Research Groups (11321503), the Knowledge